# EEG Dynamic Microstate Patterns Induced by Pulsed Wave Transcranial Photobiomodulation Therapy

**Jiangshan He[1,2], Hui Xie[1,2,3,*], Yuqiang Yang[1,2], Chunli Jia[1,2], Dan Liang[1,2], Lianghua Zhang[1,2], Xiaoyu Wang[1,2], Tianyi Luo[1,2], Zexiao Dong[1,2], Huiting Yang[1,2], Zhen Yuan [4], Mingzhe Jiang[5,*], Xueli Chen[1,2,3,5,6,*]**

[1] *Center for Biomedical-photonics and Molecular Imaging, Advanced Diagnostic-Therapy Technology and Equipment Key Laboratory of Higher Education Institutions in Shaanxi Province, School of Life Science and Technology, Xidian University, Xi'an, Shaanxi 710126, China;*

[2] *Xi'an Key Laboratory of Intelligent Sensing and Regulation of trans-Scale Life Information, School of Life Science and Technology, Xidian University, Xi'an, Shaanxi 710126, China;*

[3] *State Key Laboratory of Electromechanical Integrated Manufacturing of High-Performance Electronic Equipment, Xidian University, Xi'an, Shaanxi 710071, China;*

[4] *Centre for Cognitive and Brain Sciences, University of Macau, Macau SAR, China*

[5] *Bi-optoelectronic-integration and Medical Instrumentation Laboratory, Guangzhou Institute of Technology, Xidian University, Guangzhou, Guangdong 510555, China;*

[6] *Chongqing Institute for Brain and Intelligence, Guangyang Bay Laboratory, Chongqing, 400064, China*

* *Correspondence:* mzjiang@xidian.edu.cn (M. J.), hxie@xidian.edu.cn (H. X.), xlchen@xidian.edu.cn (X. C.)

## Abstract

### Significance

Transcranial photobiomodulation (tPBM) therapy is an emerging, non-invasive neuromodulation technique that has demonstrated considerable potential in the field of neuropsychiatric disorders. Several studies have found that pulsed wave (PW) tPBM therapy yields superior biomodulatory effects. However, its neural mechanisms are still unknown which poses a significant barrier to the development of an optimized protocol.

### Aim

This study aims to employ EEG dynamic microstate analysis combined with explainable machine learning approaches to unveil microstate dynamics and identify key features for PW tPBM.

### Approach

A randomized, single-blind study including 29 participants was conducted using a crossover design, where the PW tPBM was administered, with sham and continuous wave (CW) groups as controls. The EEG microstate analysis was utilized to investigate the relative variations in temporal parameters and brain functional connectivity. To further elucidate the dynamic activity patterns of microstates, a 10-repeat 10-fold cross-validation with nine machine learning algorithms and kernel Shapley additive explanations analysis was employed.

### Results

Statistical analysis indicated that the PW tPBM enhanced the global efficiency, local efficiency, and betweenness centrality of microstate C in brain functional networks as well as the mean durations parameter achieving a middle to large effect size, with superior effects compared to the sham and CW groups. Furthermore, the support vector machine based on the radial basis function method with kernel Shapley additive explanations analysis demonstrated the best performance with an area under the curve (AUC) reaching 0.956, and found that the 8 of top-10 microstate features related to microstate C contributed most significantly to the PW tPBM.

**Conclusions**

In conclusion, the EEG microstate analysis found that PW tPBM therapy modulates the microstate C-specific patterns in the human brain, suggesting that microstate dynamics may serve as a state-dependent biomarker for the optimization of tPBM protocol.

**Key words** transcranial photobiomodulation; pulsed wave; EEG microstate; explainable machine learning

## Introduction

Endre Mester's failed tumor eradication experiment using a ruby laser (694nm) unexpectedly unveiled the non-thermal regulatory effect of photobiomodulation (PBM) therapy in 1967, evidenced by accelerated hair growth and wound healing [1]. Since then, PBM therapy was initially predominantly investigated for its efficacy in promoting wound healing [2,3] and alleviating pain [4] among other applications. Over the past two decades, transcranial PBM (tPBM) therapy has demonstrated substantial progress in the treatment of stroke [5–7], major depression disorder [8–10], and Alzheimer's disease [11,12] as a non-invasive and safe near-infrared light-based neuromodulation technique. tPBM therapy modulates cytochrome c oxidase and activates mitochondrial metabolic pathways, subsequently triggering a cascade of downstream signaling pathways that regulate cellular physiological functions and achieve neuroprotective effects [1,13].

However, previous studies have revealed considerable heterogeneity in intervention protocols [14–16]. Consequently, to facilitate the clinical translation of tPBM therapy, it remains imperative further to elucidate the action mechanism of light intervention parameters. Among numerous parameters, the PW mode—rapid temperature pulses—can induce an increase in cell membrane capacitance, subsequently generating predictable excitatory ionic displacement currents that lead to neuronal excitation [17]. From the standpoint of improving clinical outcomes, in previous studies, compared to the CW and sham mode, the PW mode offers better biomodulatory effects in wound healing [2,3], pain [4], tissue regeneration [18–20], stroke [21–24], cognitive performance, and mental state [25,26]. From a safety standpoint, PW mode with high peak and low average power density, safely reduces cumulative thermal exposure to mitigate skin tissue injury within biological thermal tolerance thresholds [27,28]. Dong et al. reported that Alzheimer's patients with elevated brain gray matter energy deposition showed minimal decline in activities of daily living scores, suggesting a seemingly correlation between energy accumulation and functional enhancement [29].

Despite its preliminary evidence of clinical effectiveness, the neural mechanism underlying the effects of PW mode in human brain activity remains poorly understood.

To evaluate the therapeutic benefits of PW mode, it is necessary to establish evidence of neurophysiological effects by the human brain functional changes induced by the PW mode and others. To date, relevant evidence is mostly from experiments on traumatic brain injury in mice [23], a case report of a mixture of CW and PW modes [30], and Tang et al. found that the PW mode exhibited superior memory and vigilance performance, without inducing alterations in EEG spectral patterns [25]. EEG offers high temporal resolution electrophysiological evidence crucial for elucidating the neural mechanisms of tPBM therapy. Several studies have indicated that tPBM modulates electrophysiological activity and corresponding brain functional networks on the resting-state [15,16]. For instance, tPBM can enhance connectivity patterns and information transmission in the human brain [31]. And it selectively upregulates alpha-wave power and optimizes attentional resource allocation[32]. Additionally, tPBM can alter the EEG microstate of the resting brain which represents brain activation across the frontal and parietal regions [33]. However, these studies predominantly employed a single stimulation protocol and exhibited variability across studies. Therefore, data demonstrating the effects of PW tPBM on the human brain is lacking.

To investigate the neural mechanisms of PW tPBM therapy, this study proposed a randomized, single-blind, controlled, crossover design that included three 8-minute tPBM sessions, with PW, CW and sham groups, and a washout period of at least 7 days. To comprehensively investigate the impact of PW mode on resting-state brain functional activity, between-group comparisons of the relative changes relative to the baseline were conducted. Firstly, through the statistical analysis of microstate parameters, we can gain an in-depth understanding of the temporal dynamic characteristics of brain functional activities. Then, the weighted phase lag index-based dynamic microstate brain network analysis can precisely identify the collaborative working patterns among different brain regions under specific microstate, thereby revealing the functional integration mechanisms of brain regions. Finally, machine learning algorithms with the kernel shapley additive explanations (SHAP) analysis are capable of handling complex multi-dimensional features and delve into the hidden patterns and rules in microstate analysis. Given that active tPBM therapy can modulate brain functional activity at resting-state, we hypothesized that PW tPBM therapy can trigger notable activation of brain function, and such activation will be evident in multiple electrophysiological dimensions.

## Methods

### 1. Participants

A total of 33 participants (11 female) were recruited. All of them are right-handed, with a mean age of 19.52 ± 1.35 years. The inclusion criteria were age over 18 years, general mental and physical health. The exclusion criteria were as follows: including neurological or psychiatric diseases, those with a history of brain injuries or violent behavior, pregnant individuals, photosensitivity disorders, and experience of neuromodulation within the past month. The experimental procedures were approved by the Institutional Review Board of the local institutes. All participants voluntarily signed an informed consent prior to participation.

**2. Experimental setups**

This is a single-blind, crossover design study in which each participant was assigned via a random number table to take part in three tPBM sessions with varying parameters. A minimum interval of 7 days was maintained between two consecutive interventions to ensure proper washout effects. The experiment was administered using 980 nm laser (Model Aurora-A3, developed by Wuhan Jin Laser Medical Technology Co., Ltd., Hubei, China). The 980 nm has commonly been used both in preclinical and clinical research, such as stroke and wounds, demonstrating notable therapeutic effects [34–39]. In figure 1 (a), the intervention sites were positioned on the Fp2 site, as determined by the EEG 10/20 system. In figure 1 (b), the measured uniform laser beam has an area of 12.57 $cm^2$ and an average power density of 238.85 $mW/cm^2$, resulting in a total energy of ~1440 J ($12.57\ \mathrm{cm}^2 \times 238.85\mathrm{mW/cm}^2 \times 480\mathrm{s} \approx 1440\ \mathrm{J}$). Each experimental paradigm consists of three phases in figure 1 (c), a 5-minute pre-stimulation closed-eyes resting-state, followed by an 8-minute tPBM stimulation period, and a 5-minute post-stimulation closed-eyes resting-state. Checking for EEG quality by the subject took $260.16 \pm 76.84$ s after tPBM stimulation. During the tPBM stimulation, participants were instructed to wear laser protective eyewear, and keep their eyes closed and were instructed to keep relaxed during the experiment.

We simulated the energy deposition profiles to quantify how different tPBM sessions, namely CW and PW mode, affected targeted brain regions for this study. Finite element method simulations were conducted using *MCX-1.2* and *MATLAB R2023a* to model the energy deposition induced by tPBM. The Colin27 standard brain template and brain optical parameters has been used to generate simulations in previous studies [40,41]. To delineate distinct cortical functional regions, we employed the Brainnetome atlas [42] for parcellation and segmentation of the Colin27 standard brain template. Simulation results demonstrate that: the A10m_R in the superior frontal gyrus (SFG), along with A46_R and A10l_R in the middle frontal gyrus (MFG), constitute the primary brain regions with significant energy accumulation (Supplementary figure S1 and S2). Overall, the peak energy accumulation in PW mode is about twice that of CW mode, and their average energy accumulations are comparable, in figure 1 (d) and (e).

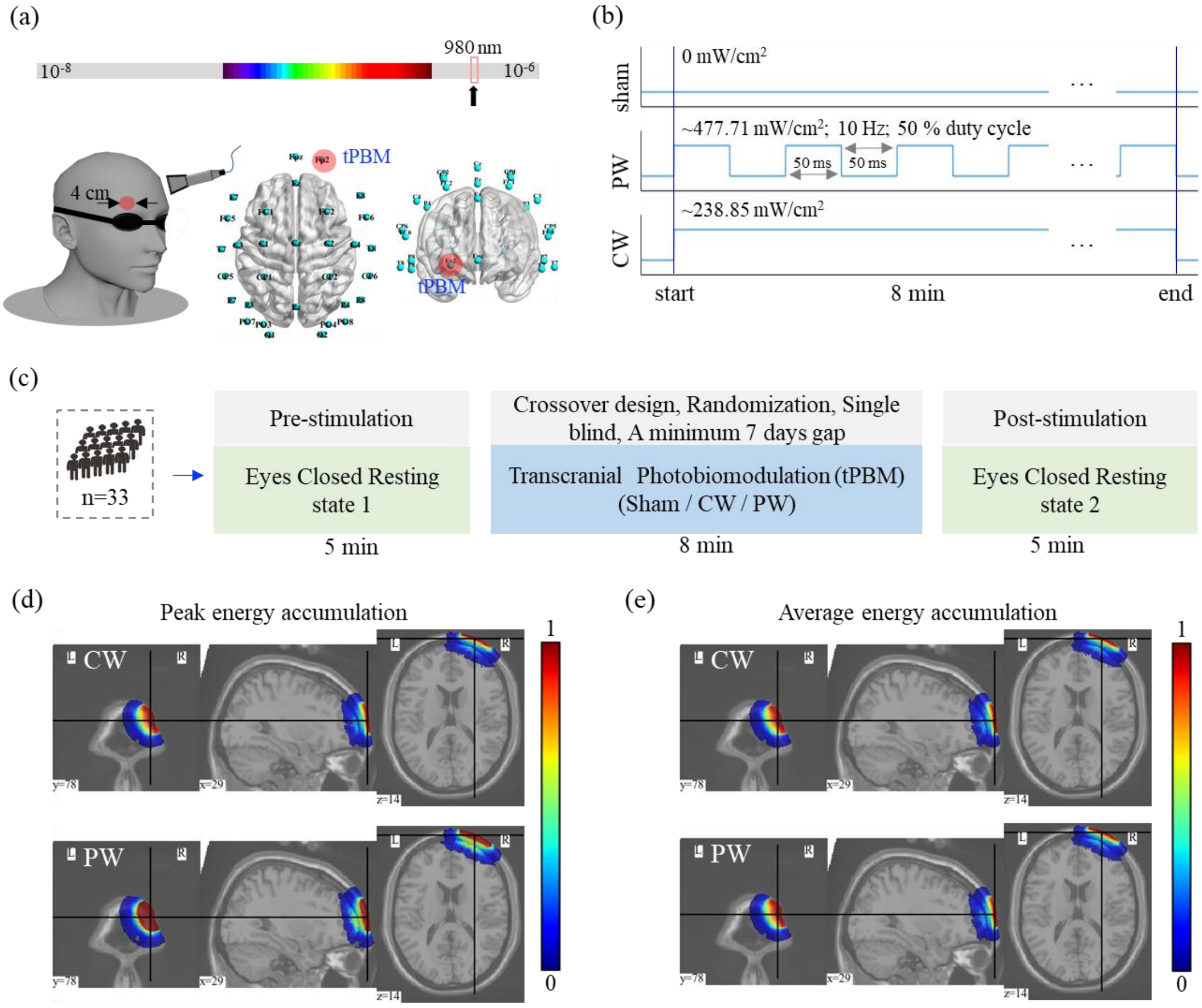

Figure 1. Experimental configurations for PW tPBM investigation. (a) the EEG electrode montage and tPBM stimulation target region, (b) protocol schematic of tPBM parameter settings, such as wavelength, irradiance, pulse frequency and so on, (c) Temporal arrangement of tPBM interventions session, the spatial distribution of peak (d) and average (e) energy accumulation in Colin27 standard brain template, respectively.

## 3. Data collection and preprocessing

EEG data were collected using the Emotiv Flex2 system under the eye-closed resting-state condition. The channel location was according to the EEG 10/20 system, and the bilateral earlobes served as the reference electrodes. To ensure high-quality signal acquisition, all electrodes were prepared with Greentek GT5 conductive gel to reduce impedance before the data recording. Data preprocessing was performed using the *MNE-1.7.0*. The EEG data were sampled at 128 Hz and processed with a 2-20 Hz bandpass filter [43,44], and REST re-referencing was subsequently applied [45]. To mitigate physiological artifacts such as ocular, cardiac, and muscular noise, independent component analysis (ICA) was employed. Artifact-contaminated segments exceeding ±150 μV in amplitude were manually identified and removed. Finally, 29 participants were retained for analysis, while the other 4 participants (1 female) were excluded due to dropout, absence of alpha waves, motion artifacts, or inadequate signal quality.

## 4. EEG microstate analysis

EEG microstate analysis serves as atoms of thought, and brain activity processes can be described by a series of alternating, stable, and limited microstate sequences in the resting-state scalp electric field distribution[46]. Various types of EEG microstate classes are determined by corresponding specific resting-state networks[47]. Microstate analysis [46] was performed using the *Pycrostates-0.7.0*. For each participant, the Global Field Power (GFP) peaks were extracted, and a random subset of 5000 GFP peaks was selected. The modified k-means microstate clustering algorithm was applied, with a smoothing factor of 10, a sliding window width of 5, and a minimum duration of 3 [43,44]. Then, four microstate classes were identified, and the microstate sequences were derived for each participant during a 5-minute eyes-closed resting-state.

### a) Microstate parameter analysis

The microstate parameters were computed based on the microstate sequences, including mean durations, occurrences per second, time coverage and transition probability [48]. Mean durations reflect the average temporal stability of each microstate. Occurrences per second quantify the frequency of microstate reappearances. Time coverage represents the proportion of total recording time occupied by each microstate. Transition probability (TP) $P\{X(m+1)=j \mid X(m)=i\}$, where $i,j \in I$ and $I$ is the microstate space, comprising four microstate classes. TP is referred to as the one-step transition probability of the Markov chain {X(n), n≥0} from state *i* to state *j* at time m. Then the relative microstate temporal parameters $\Delta MS_i$ were computed and normalized to the pre-stimulation period:

$$\Delta MS_i = \left(MS_{i,post} - MS_{i,pre}\right) / MS_{i,pre} \tag{1}$$

Where *i* represents the mean durations, occurrences per second, time coverage, and transition probability.

### b) Dynamics microstate brain functional network analysis

Conventional brain functional connectivity is typically assessed using fixed-window approaches, whereas EEG microstate provide a temporal dynamics perspective for investigating brain functional networks [49]. The weighted phase lag index (wPLI) quantifies the phase synchronization between different brain areas, which is a robust and widely used method for functional connectivity estimation [50,51], and it was defined as follows:

$$wPLI = \left| \frac{\sum_{t=1}^{n} \left| imag\left(S_{xy,t}\right) \right| \operatorname{sgn}\left( imag\left(S_{xy,t}\right)\right)}{\sum_{t=1}^{n} \left| imag\left(S_{xy,t}\right) \right|} \right| \tag{2}$$

where the complex cross-spectral density $S_{xy,t}$ characterizes the phase relationship between two real-valued signals $x(t)$ and $y(t)$ within a time window *t*. EEG microstate provide insights into the rapid temporal dynamics of neural activity [49]. Therefore, the microstate-based time window strategy is employed, where *t* denotes a non-overlapping temporal segment corresponding to a specific microstate class, rather than a conventional fixed window. And its value ranged from 0 to 1. A low value indicates the weak synchronization between signals from two brain regions, whereas a high value indicates the enhanced synchronization.

Then, the graph theory was applied to quantify the topological changes in brain networks via pairwise functional connectivity between channels [52,53]. The graph theory metrics include the efficiency metrics (global efficiency and local efficiency) and nodal graphical metrics (betweenness centrality) [54]. These metrics were computed across 19 sparsity thresholds (5%–95%, in 5% increments) to evaluate their robustness under varying network densities. Local efficiency and global efficiency metrics are used to assess the average information transfer efficiency at the subgraph and global levels, respectively [55]. Higher efficiency values reflect more optimal information propagation within the network. The betweenness centrality quantifies a node's role as a bridge between nodes, defined as the fraction of shortest paths connecting all node pairs that traverse the given node [54]. A node with a high betweenness centrality score indicates that it serves as a critical conduit for information flow.

**c) Explainable machine learning analysis**

In this study, nine machine learning algorithms were employed to evaluate performance in the classification using *Scikit-learn 1.5.2*. A min-max normalization process was applied to the dataset to ensure more stable model convergence. For the performance evaluation, we employed the K-nearest neighbors (KNN), the linear support vector machine (linear SVM), the support vector machine based on radial basis function (RBF SVM), the Gaussian process, the decision tree, the random forest, the Adaboost, the Navie bayes, and the quadratic discriminant analysis (QDA) algorithms for three-class classification. The dataset comprised three groups, each containing 29 samples with 44 microstate features. These included 24 microstate parameters (mean durations, occurrences per second, time coverage, and transition probability for each of the four microstates), and 20 network metrics at a 50 % sparsity level (global efficiency, local efficiency, and betweenness centrality for each of the four microstates). A 10-repeats 10-fold cross-validation on the microstate features dataset. Performance was assessed using the area under the curve (AUC), accuracy, precision, recall, and F1-score, averaged across all test folds from the 10 repeats. To further elucidate the comprehensive alterations of EEG microstate features under the three tPBM sessions, the kernel SHAP analysis [56,57] was utilized to investigate temporal discrepancies in microstate parameters. The kernel SHAP specifies the explanation as follows:

$$g(y')=\phi_0+\sum_{j=1}^{M}\phi_j y_j' \tag{3}$$

Where the explanation model is denoted by $g(y')$, $\emptyset_j$ is the effect of each feature, and *M* is the number of input microstate temporal parameters $y_j'$.

## 5. Statistical methods

Statistical analyses were performed using *Scipy-1.14.1* and *Pingouin-0.5.* The normality of distribution is assessed by the *normal* test function. For normally distributed data, the one-way ANOVA was used to compare the three groups, followed by a pairwise Tukey-HSD post-hoc test. For non-normally distributed data, the non-parametric Kruskal-Wallis H-test was used to compare the three groups, followed by a pairwise Games-Howell post-hoc test. Cohen's d was calculated as a measure of the effect size for pairwise comparisons. The Benjamini Hochberg FDR correction is used to control multiple comparison problems, with statistical significance defined as $p < 0.05$. Data are presented as mean ± standard deviation.

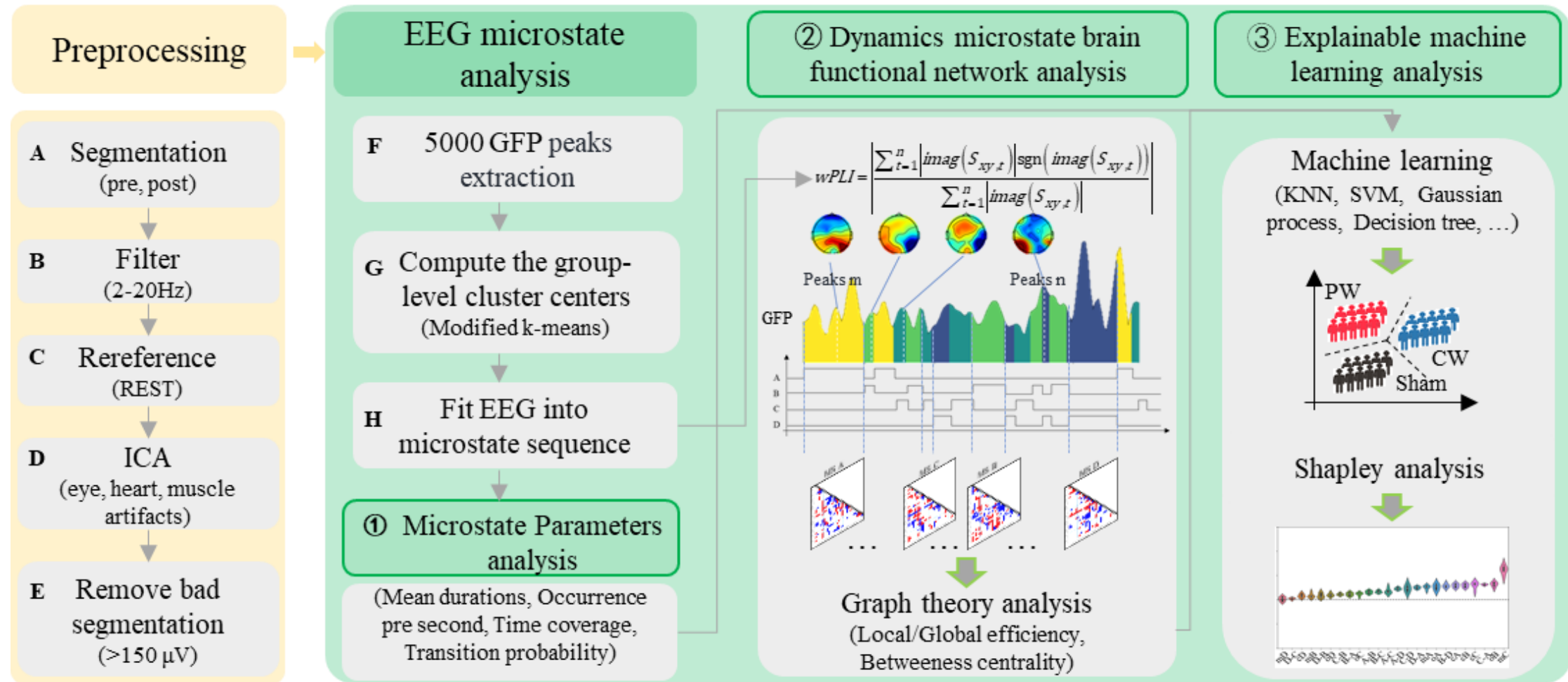

Figure 2. Flowchart of EEG data processing which consists of preprocessing and EEG microstate analysis.

## Results

### 1. Identification of EEG microstate classes

The global explained variance (GEV) score was 0.701 ± 0.017, indicating the model reconstruction effectiveness. Microstate topographic maps across the three groups aligned with prior findings [58,59]. In figure 3(a), microstate A (MS A) exhibited a right frontal-to-left posterior configuration, microstate B (MS B) exhibited a left frontal-to-right posterior configuration, microstate C (MS C) presented a fronto-central configuration, and microstate D (MS D) displayed a centro-parietal maximum. These microstate classes represented recurrent patterns of EEG activity that were consistently observed across groups in figure 3(b).

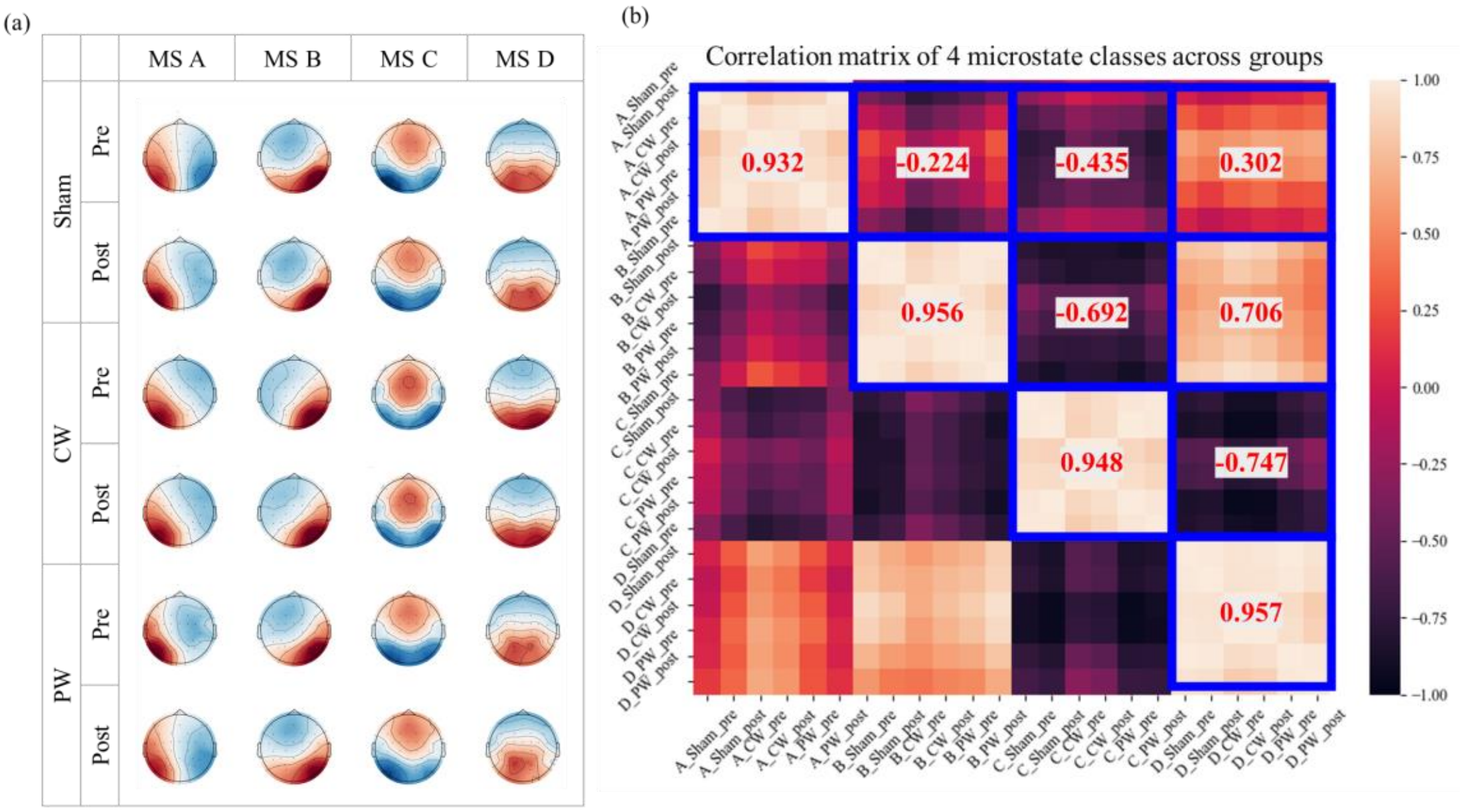

Figure. 3 Topographic maps in the sham, CW, and PW groups. (a) The 4 microstate classes (microstate A-D) across groups before (Pre) and after (Post) the tPBM therapy, (b) correlation matrix.

## 2. Microstate parameters analysis

The mean duration as shown in figure 4 (a) and table I, there existed significant differences among the three groups for MS A (H-statistic=10.878, $p=4.345e^{-3}$) and MS C (H-statistic=16.937, $p=2.100e^{-4}$). Post-hoc analyses revealed that the mean duration of MS A in the PW group was significantly lower than in the CW group (T=3.459, p=0.009, Cohen's d=0.908). For MS C, the mean duration in the PW group was significantly higher than in the sham (T=3.173, p=0.014, Cohen's d=0.833) and CW (T=3.265, p=0.014, Cohen's d=0.857) groups.

As illustrated in the microstate temporal parameters above, compared to the sham and CW groups, the PW group exhibited a significantly reduced mean duration of MS A and a significantly prolonged mean duration of MS C. These findings suggest that PW conditions may suppress MS A while enhancing MS C. While parametric trends in the CW group were intermediate between sham and PW groups, no statistically significant differences were observed between CW and sham groups.

The transition probability as shown in figure 4 (d) and table II, there existed significant difference among the three groups for $TP_{A\text{-}C}$ (H-statistic=15.331, $p=4.688e^{-4}$), and post-hoc analyses indicated that $TP_{A\text{-}C}$ in the PW group was significantly higher than in the sham (T=3.197, p=0.020, Cohen's d=0.840). $TP_{A\text{-}D}$ (H-statistic=10.754, $p=4.621e^{-3}$) showed a significant difference among the three groups, and post-hoc analyses indicated that $T_{A\text{-}D}$ in the PW group was significantly lower than in the sham (T=3.016, p=0.032, Cohen's d=0.792). $TP_{B\text{-}C}$ (H-statistic=23.132, $p=9.482e^{-6}$) showed a significant difference among three groups, and post-hoc analyses indicated that $TP_{B\text{-}C}$ in the PW group was significantly higher than in the sham (T=4.084, p=0.001, Cohen's d=1.072) and CW (T=2.727, p=0.032, Cohen's d=0.716) group, $T_{B\text{-}C}$ in the sham group was significantly lower than in the CW (T=2.612, p=0.032, Cohen's

d=0.686) group. $TP_{B-D}$ (F-statistic=7.289, $p=1.205e^{-3}$) showed a significant difference among the three groups, and post-hoc analyses indicated that $TP_{B-D}$ in the PW group was significantly lower than in the sham (T=3.623, p=0.004, Cohen's d=0.940) and CW (T=2.855, p=0.022, Cohen's d=0.834) group. $TP_{D-C}$ (H-statistic=13.392, $p=1.236e^{-3}$) showed a significant difference among the three groups, and post-hoc analyses indicated that $TP_{D-C}$ in the sham group was significantly lower than in the CW (T=3.304, p=0.016, Cohen's d=0.868).

As evidenced by the alteration of transition probability, there is a statistically significant difference between sham and CW group, the PW group exhibited significantly higher values for $TP_{A-C}$ and $TP_{B-C}$ while demonstrating lower values for $TP_{A-D}$ and $TP_{B-D}$. The sham group, serving as control, showed lower values for $TP_{A-C}$, $TP_{B-C}$, and $TP_{D-C}$ compared to both the CW and PW groups. The CW group consistently fell between the PW and sham groups in $TP_{B-C}$. The significant differences between CW and PW groups in $TP_{B-D}$ where CW had higher parameters compared to PW.

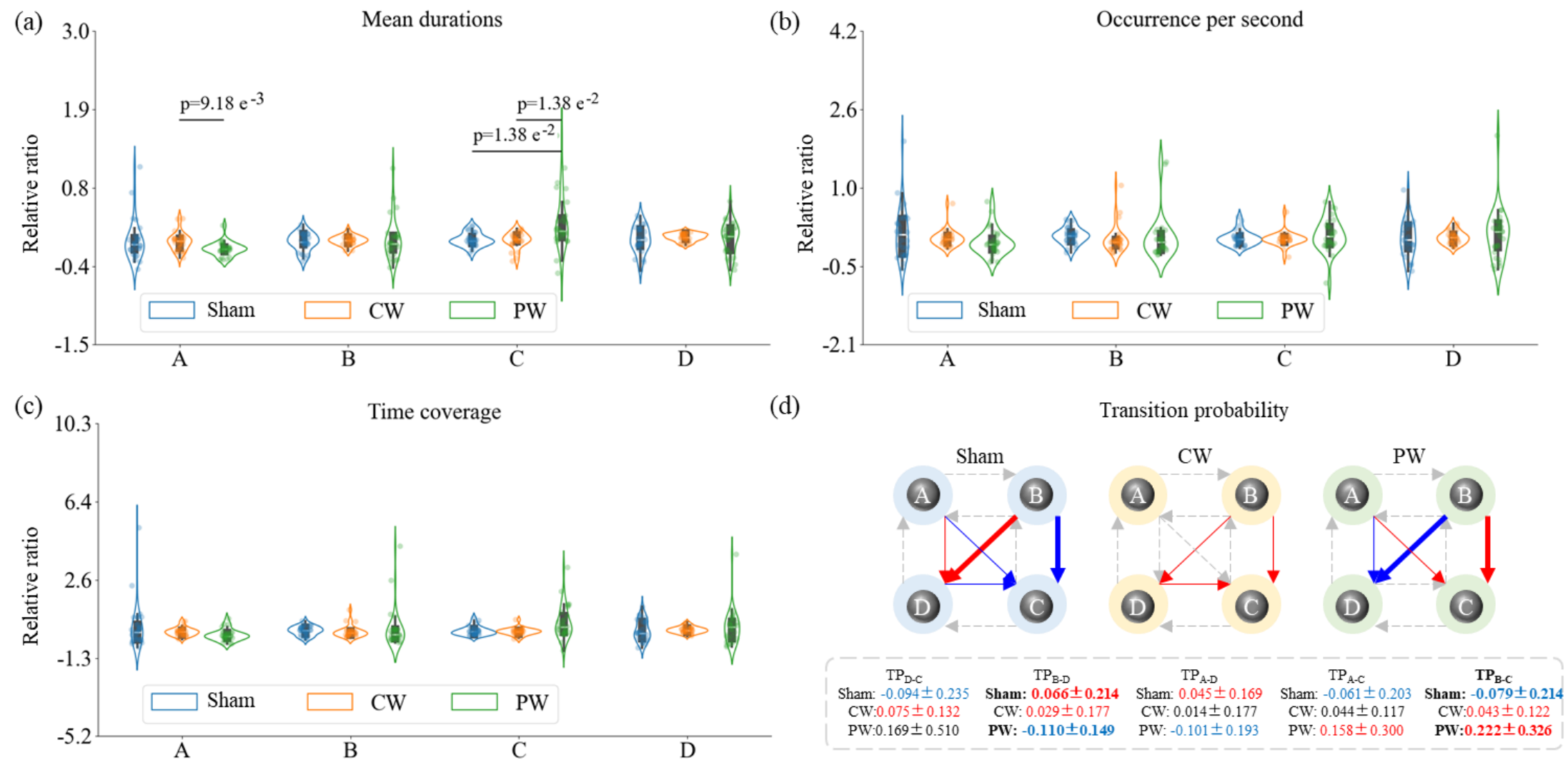

Figure 4. Comparison of the EEG microstate parameters among sham, CW, and PW groups. (a) Mean durations; (b) Occurrence pre second; (c) Time coverage; (d) Transition probability. Line thickness represents the magnitude of relative changes ratio, the red line indicates an increase and the blue line indicates a decrease trend.

**Table I.** Comparison of EEG microstate parameters between sham, CW and PW groups.

| Microstate parameters | | Mean ± standard deviation | | | statistic (H/F) | P<0.05 | Post-hoc (T-value, $P_{FDR_bh}$ <0.05, Cohen's d) | | |
|---|---|---|---|---|---|---|---|---|---|
| | | sham | CW | PW | | | CW vs. sham | CW vs. PW | PW vs. sham |
| Mean durations | A | -0.008±0.294 | -0.025±0.128 | -0.131±0.099 | **10.878** | **$4.345e^{-3}$** | -0.279, 0.958, -0.073 | **3.459, 0.009, 0.908** | -2.089, 0.161, -0.548 |
| | B | -0.027±0.131 | -0.006±0.070 | 0.008±0.297 | 0.863 | $6.494e^{-1}$ | - | - | - |
| | C | 0.007±0.099 | -0.002±0.113 | 0.254±0.400 | **16.937** | **$2.100e^{-4}$** | -0.329, 0.942, -0.086 | **-3.265, 0.014, -0.857** | **3.173, 0.014, 0.833** |
| | D | 0.007±0.191 | 0.044±0.056 | 0.032±0.225 | 0.695 | $7.063e^{-1}$ | - | - | - |
| Occurrence | A | 0.142±0.510 | 0.035±0.163 | -0.040±0.263 | 4.004 | $1.350e^{-1}$ | - | - | - |
| | B | 0.084±0.163 | 0.013±0.266 | 0.051±0.435 | **8.044** | **$1.791e^{-2}$** | -1.209, 0.923, -0.317 | -0.393, 0.923, -0.103 | -0.381, 0.923, -0.100 |
| | C | 0.040±0.147 | 0.019±0.138 | 0.101±0.322 | 0.905 | $6.362e^{-1}$ | - | - | - |
| | D | 0.039±0.362 | 0.050±0.125 | 0.119±0.502 | 0.618 | $7.342e^{-1}$ | - | - | - |
| Time coverage | A | 0.263±1.105 | 0.011±0.207 | -0.155±0.281 | **7.189** | **$2.747e^{-2}$** | -1.185, 0.471, -0.311 | 2.526, 0.115, 0.663 | -1.941, 0.216, -0.510 |
| | B | 0.059±0.227 | 0.008±0.280 | 0.157±0.963 | 4.163 | $1.248e^{-1}$ | - | - | - |
| | C | 0.054±0.222 | 0.016±0.176 | 0.469±0.807 | **8.406** | **$1.495e^{-2}$** | -0.702, 0.764, -0.184 | -2.900, 0.051, -0.762 | 2.624, 0.051, 0.689 |
| | D | 0.090±0.506 | 0.098±0.157 | 0.245±0.849 | 0.571 | $7.516e^{-1}$ | - | - | - |

**Table II.** Comparison of EEG microstate transition probability parameters between sham, CW and PW groups.

| TP | | Mean ± standard deviation | | | | | Post-hoc (T-value, $P_{FDR_bh}$<0.05, Cohen's d) | | |
|---|---|---|---|---|---|---|---|---|---|
| From | To | sham | CW | PW | statistic (H/F) | P<0.05 | CW vs. sham | CW vs. PW | PW vs. sham |
| A | B | 0.081±0.217 | -0.010±0.160 | 0.033±0.296 | 2.893 | $2.354e^{-1}$ | - | - | - |
| | C | -0.061±0.203 | 0.044±0.117 | 0.158±0.300 | **15.331** | **$4.688e^{-4}$** | 2.374, 0.084, 0.623 | -1.872, 0.161, -0.492 | **3.197, 0.020, 0.840** |
| | D | 0.045±0.169 | 0.014±0.177 | -0.101±0.193 | **10.754** | **$4.621e^{-3}$** | -0.681, 0.775, -0.179 | 2.320, 0.092, 0.609 | **-3.016, 0.032, -0.792** |
| B | A | 0.119±0.353 | -0.015±0.210 | 0.022±0.271 | 2.319 | $3.136e^{-1}$ | - | - | - |
| | C | -0.079±0.214 | 0.043±0.122 | 0.222±0.326 | **23.132** | **$9.482e^{-6}$** | **2.612, 0.032, 0.686** | **-2.727, 0.032, -0.716** | **4.084, 0.001, 1.072** |
| | D | 0.066±0.214 | 0.029±0.177 | -0.110±0.149 | **(F)7.289** | **$1.205e^{-3}$** | -0.768, 0.724, -0.187 | **2.855, 0.022, 0.834** | **-3.623, 0.004, -0.940** |
| C | A | 0.056±0.212 | 0.019±0.317 | 0.010±0.268 | 2.253 | $3.242e^{-1}$ | - | - | - |
| | B | 0.044±0.181 | 0.025±0.147 | 0.164±0.440 | 3.004 | $2.226e^{-1}$ | - | - | - |
| | D | -0.038±0.189 | 0.071±0.254 | -0.083±0.221 | **6.809** | **$3.323e^{-2}$** | 1.824, 0.258, 0.479 | 2.414, 0.148, 0.634 | -0.810, 0.699, -0.213 |
| D | A | 0.128±0.323 | -0.017±0.274 | 0.006±0.327 | **8.528** | **$1.406e^{-2}$** | -1.819, 0.511, -0.478 | -0.279, 0.958, -0.073 | -1.415, 0.511, -0.371 |
| | B | 0.089±0.206 | 0.017±0.172 | 0.012±0.240 | 5.689 | $5.817e^{-2}$ | - | - | - |
| | C | -0.094±0.235 | 0.075±0.132 | 0.169±0.510 | **13.392** | **$1.236e^{-3}$** | **3.304, 0.016, 0.868** | -0.953, 0.611, -0.250 | 2.479, 0.068, 0.651 |

## 3. Dynamic microstate brain functional network analysis

To investigate tPBM-induced alterations in functional brain networks, we quantified microstates-based functional connectivity for the four microstate classes in figure 5. For each participant, the 30×30 connectivity matrix was used by quantifying the wPLI based on Eq. (2) under both the tPBM and sham conditions. Following the Benjamini Hochberg FDR correction, suprathreshold edges were identified, and a Cohen's d score of 0.5 (medium ES) was selected as the critical value. Microstates-based functional connectivity analysis revealed a reciprocal relationship between PW and sham groups, while CW group exhibited an intermediate alteration in connectivity strength.

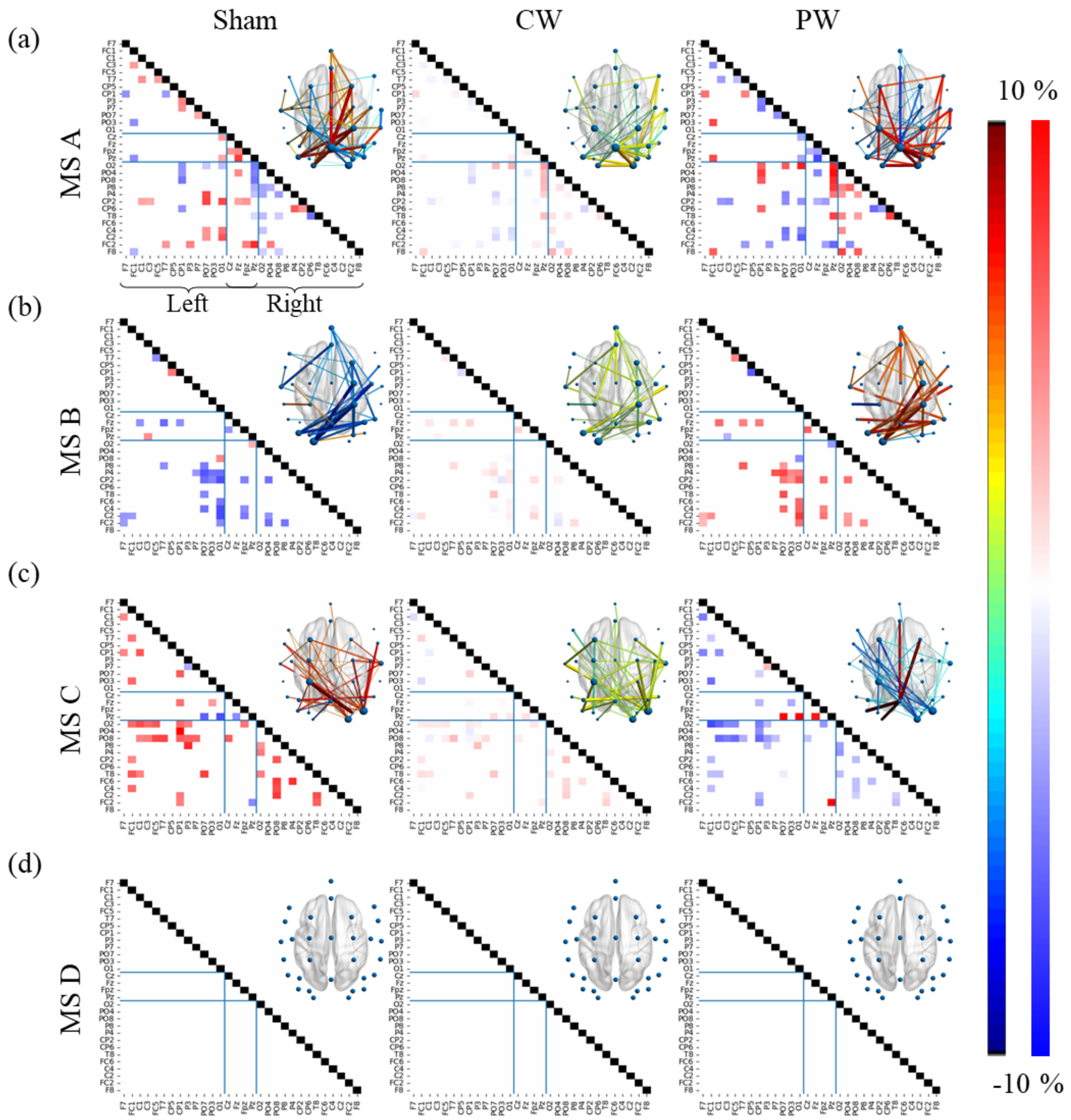

Figure 5. Microstate-specific brain functional network of MS A(a), MS B (b), MS C (c), MS D (d) in the sham, CW, and PW groups, averaged over n=29 young adults. Topographic distribution of thirty EEG electrodes partitioned into left, right, and middle regions (blue lines demarcate functional groupings).

To investigate PW mode induced alterations in dynamic microstate brain functional network, graph theory was employed to quantified microstate-based brain functional connectivity. For the local efficiency (LE) metric, in figure 6 (a), synchronization in the CW and PW groups exceeded that of the sham group at sparsity levels of 15% and 25%-55% for MS B. Conversely, for MS C, PW group synchronization was significantly lower than CW and sham groups at sparsity levels of 25%-45%, 60%, and 65%. Similarly, for global efficiency (GE) in figure 6 (b), CW and PW groups demonstrated higher synchronization than the sham group across sparsity levels of 15%-55% for MS B. For MS C, PW group synchronization was reduced compared to CW and sham groups at sparsity levels of 30%-45% and 65%-80%. In figure 6 (c), betweenness centrality (BC) analysis revealed significantly lower values in CW and PW groups relative to sham at sparsity levels of 15%-50% for MS B. For MS C, PW group BC was elevated at 30%-45% sparsity but reduced at 70%-80% sparsity. Comparable trends were observed for the left hemispheric BC in figure 6(d) and right hemispheres in figure 6(e). In summary, tPBM enhanced LE and GE metrics for MS B in both CW and PW groups, whereas only PW group LE and GE for MS C were diminished. BC was reduced in MS B under tPBM, with MS C exhibiting elevated low-threshold BC in the PW group but attenuated high-threshold BC.

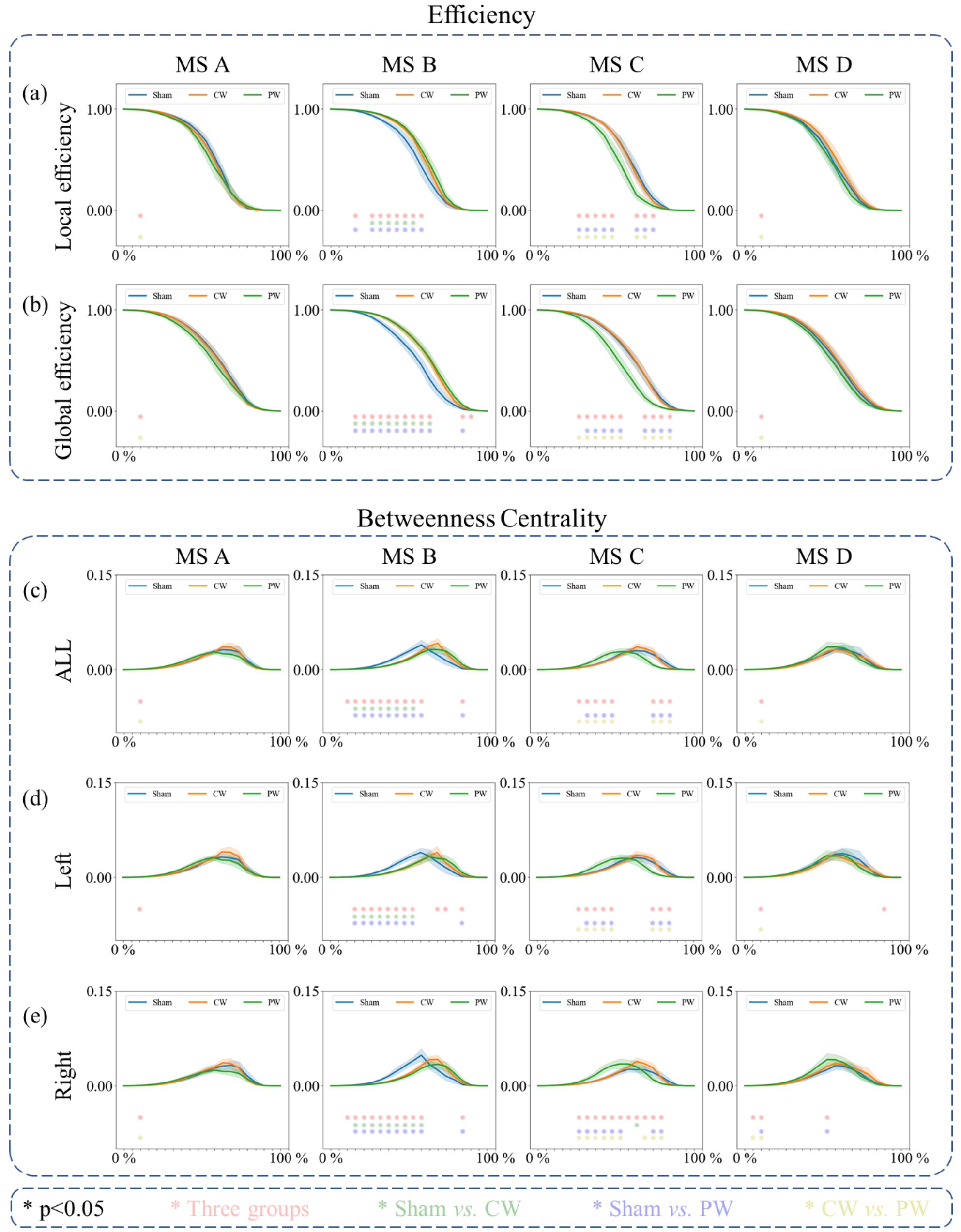

Figure 6. Quantitative analysis of brain synchronization based on graph theory at 2-20 Hz frequency band. The global efficiency (a), the local efficiency (b), the betweenness centrality of the whole brain (c), the left hemisphere (d), and the right hemisphere (e). The red stars mark at different sparsity levels represent for the corresponding metric has statistical significance ($p<0.05$) among three groups by the one-way ANOVA or the non-parametric Kruskal-Wallis H-test. For the post-hoc test, the green stars for the sham and CW groups, the blue stars for the sham and PW groups, and the yellow stars for the CW and PW groups. The shadow regions represent for the standard deviation.

## 4. Machine learning algorithms with SHAP analysis

We employed nine machine learning algorithms to investigate the discrepancies between tPBM sessions. In figure 7, the RBF SVM, Gaussian process and KNN algorithms have good classification accuracy exceeding 80%. In table III, the RBF

SVM exhibited the highest performance, with an F1-score of 85.692%, accuracy of 85.595%, specificity of 86.020%, and sensitivity of 85.595%. The AUC value for the RBF SVM was 0.956, close to 1. This indicates that microstate features effectively serve as neuroelectrophysiological biomarkers for distinguishing different tPBM sessions.

To further reveal the comprehensive alterations of microstate temporal parameters, table IV illustrates the ranked microstate features from the top-three algorithms. Furthermore, the importance ranking of microstate features with 8 of top-10 features specific to MS C for PW mode. The complete ranking of parameters is provided in Supplementary table SI and figure S8-10. In the PW group, particularly with regard to mean durations and time coverage of MS C accounted for 4.118 % and 3.591 %, respectively. MS A and MS C related parameters in the CW group also played a significant role, such as mean durations of MS C (2.133 %) and MS A (2.109 %). In the sham group, the microstate parameters importance exceeded 2%, both the $TP_{A\text{-}C}$ (2.33 %), the mean durations of MS C (2.063 %), and the time coverage of MS A (2.012 %). In summary, explainable machine learning models demonstrate that tPBM sessions specifically modulate microstate parameters related to MS A and MS C, particularly MS C in the PW group. These findings also provide comprehensive microstate evidence supporting the neuromechanism of tPBM.

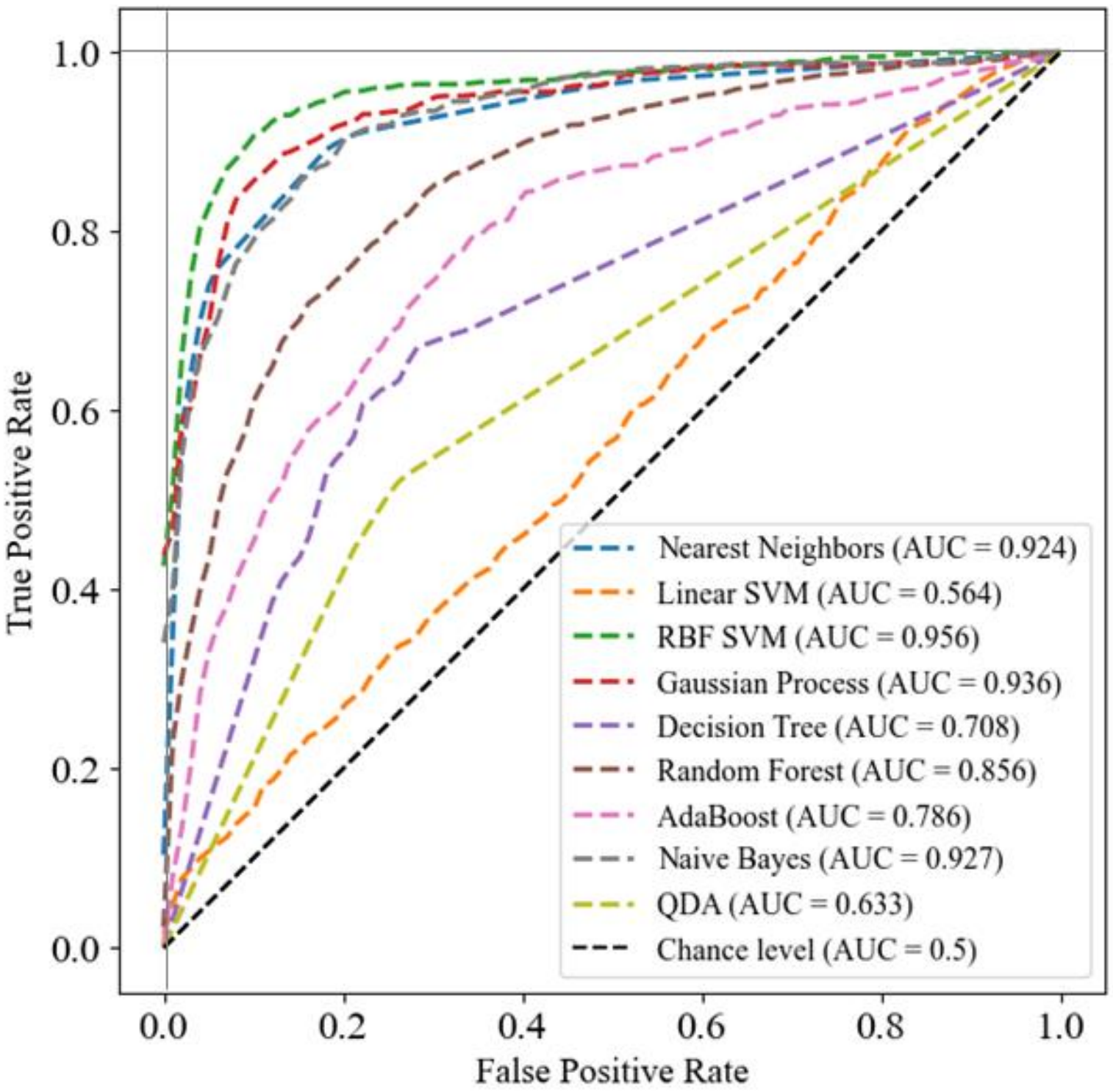

Figure 7. The AUC curve of nine algorithms for tPBM sessions classification.

**Table III. Comparison of different algorithms for tPBM classification.**

| | F1 score | Accuracy | Specificity | Sensitivity |
|---|---|---|---|---|
| KNN | 81.826% | 81.905% | 82.045% | 81.905% |
| Linear SVM | 20.492% | 20.714% | 20.496% | 20.714% |
| **RBF SVM** | **85.692%** | **85.595%** | **86.020%** | **85.595%** |
| Gaussian Process | 83.129% | 83.333% | 83.370% | 83.333% |
| Decision Tree | 60.487% | 60.357% | 60.694% | 60.357% |
| Random Forest | 70.863% | 70.833% | 70.902% | 70.833% |
| AdaBoost | 62.832% | 62.500% | 65.976% | 62.500% |
| Naive Bayes | 79.876% | 79.762% | 80.227% | 79.762% |
| QDA | 51.048% | 51.071% | 51.157% | 51.071% |

**Table IV. The importance of top-10 microstate features of the KNN, RBF SVM, and Gaussian process algorithms.**

| No. | Sham | | CW | | PW | | Average | |
|---|---|---|---|---|---|---|---|---|
| | Importance | Feature | Importance | Feature | Importance | Feature | Importance | Feature |
| 1 | **2.33 %** | **A-C** | **2.133 %** | **mC** | **4.118 %** | **mC** | **2.771 %** | **mC** |
| 2 | **2.063 %** | **mC** | **2.109 %** | **mA** | **3.591 %** | **cC** | **2.445 %** | **cC** |
| 3 | **2.012 %** | **cA** | **1.874 %** | **cC** | **2.705 %** | **GE_C** | **1.944 %** | **A-C** |
| 4 | **1.963 %** | **D-A** | **1.795 %** | **D-A** | **2.57 %** | **A-C** | **1.819 %** | **GE_C** |
| 5 | **1.939 %** | **BC_L_B** | **1.718 %** | **C-D** | **2.213 %** | **BC_R_C** | **1.694 %** | **mA** |
| 6 | **1.871 %** | **GE_B** | 1.658 % | BC_L_A | **2.075 %** | **mA** | **1.642 %** | **D-A** |
| 7 | **1.868 %** | **cC** | **1.631 %** | **GE_C** | **1.942 %** | **B-C** | **1.575 %** | **BC_R_C** |
| 8 | **1.73 %** | **BC_B** | **1.403 %** | **GE_B** | **1.784 %** | **BC_L_C** | **1.551 %** | **cA** |
| 9 | **1.521 %** | **D-C** | **1.274 %** | **cA** | **1.591 %** | **LE_C** | **1.394 %** | **B-C** |
| 10 | **1.462 %** | **B-C** | 1.216 % | GE_A | 1.581 % | BC_L_A | 1.356 % | BC_L_A |

*mX, the mean durations of MS X; cX, the time coverage of MS X; oX, the occurrences per second of MS X; X-X, the transition probability from X to X, or $TP_{X-X}$. GE_X, the global efficiency of MS X. LE_X, the local efficiency of MS X. BC_X / BC_L_X / BC_R_X, the betweenness centrality of MS X in whole/left/right hemisphere. Bold microstate features represent the statistically significant in the comparison after tPBM session.

## Discussions

This study adopted a randomized, single-blind, crossover design to explore the specific neurophysiological alterations induced by PW tPBM therapy using the 980 nm laser. The increased mean durations of MS C, $TP_{A-C}$ and $TP_{B-C}$, and enhanced the efficiency and capacity of brain functional connectivity networks in MS C of PW mode. Furthermore, explainable machine learning confirms MS C-related features as critical discriminative features. These findings provide important neurophysiological evidence for the PW mode-based tPBM-induced changes in brain activation at the resting-state.

Compared to the sham and CW groups, the analysis of microstate parameters (figure 4, table I and II) revealed that the PW group exclusively exhibited: (1) prolonged mean duration of MS C; (2) increased $TP_{A-C}$ coupled with decreased $TP_{A-D}$; and (3) increased $TP_{B-C}$ accompanied by reduced $TP_{B-D}$. While the CW group demonstrated a comparable trend, this did not reach statistical significance. In a previous study, Truong et al. [33] report the microstate-based neural mechanism of the 8 minutes 1064 nm right prefrontal tPBM has two aspects: (1) a significant increase in the occurrence of MS A and MS C and a significant decrease in the time coverage of MS D during 4 or 8 minutes CW mode-based tPBM, (2) an increase in the $TP_{A\text{-}C}$ during 4 or 8 minutes CW mode-based tPBM. In our study similarly identifies MS A and MS C as responsive to PW mode, but albeit with differing microstate parameters after tPBM. Simultaneous EEG-fMRI study [47] revealed that MS A is related to the auditory network, MS B is related to the visual network, MS C (canonical microstate D) is correlates with the attention network, and MS D (canonical microstate C) is linked to the salience network. In light of this evidence, it is likely that 980 nm right prefrontal PW tPBM facilitates the activation of the brain network involved in reflexive aspects of attention (attention network, the increased mean duration of MS C), and increases the focus switching and reorientation of network resources from the auditory/visual network to attentional network (the increased $TP_{A\text{-}C}$ and $TP_{B\text{-}C}$) instead of the information exchange from the auditory or visual networks to the salience network (the decreased $TP_{A\text{-}D}$ and $TP_{B\text{-}D}$).

Characterizing topological properties of microstate dynamic functional networks provides insights into the dynamic reorganization of the brain under PW mode. The results in figure 6 demonstrate that PW mode modulates MS C across a wide sparsity range. Specifically, PW mode reduces local and global efficiency, while increasing betweenness centrality at low thresholds and decreases it at high thresholds. This pattern may indicate that PW mode enhances the complexity of connectivity structures. Diminished betweenness centrality at higher sparsity thresholds and increased betweenness centrality at lower sparsity levels, the intervention promotes node interactions and information transfer, redistributing network traffic more uniformly across nodes. However, reciprocal patterns of local efficiency, global efficiency, and betweenness centrality were observed in MS-B, combined with the more transitions between MS B and MS C (the increased $TP_{B\text{-}C}$), the altered temporal dynamics of microstate-based functional networks in opposite directions likely reflect a compensatory mechanism. Previous studies have reported functional network alterations induced by tPBM at different EEG frequency bands. For instance, 800 nm tPBM applied to the right hemisphere increased the clustering coefficient and small world coefficient in the beta band while decreasing global efficiency [31]. 850 nm tPBM over the right hemisphere reduces right local processing and disrupts network synchronizability in the delta band [60], with comparable effects observed using 1064 nm tPBM [32]. 810 nm 40 Hz tPBM targeting the default mode network enhanced the efficiency of alpha and gamma brain networks [61]. In this study, we observed that graph theory metrics across different frequency bands were primarily concentrated in the alpha band for MS B. In contrast, the interventional effects of MS C appeared to be the cumulative result across multi-frequency bands (Supplementary figures S4-6). Furthermore, no statistically significant changes in EEG power topological map were detected (Supplementary figure S3), contrasting with previous reports. These findings highlight the specificity of the neural mechanisms with 980 nm PW mode-based tPBM applied to the right prefrontal region, but due to the limitations of the methodological differences, these differences must be interpreted with caution.

Explainable machine learning methods provides cognitive neuroscience researchers with learnable analytical tools. Top-three machine learning models identified significant microstate parameters under different tPBM sessions (table III and IV). The most notable change was observed in MS C which may be the key reason for the effectiveness of the PW mode. To enhance the responsiveness of tPBM therapy in psychiatric disorders, integrating machine learning with microstate-based brain function analysis techniques has the potential to serve as a promising technique for improving the personalization and real-time modulation of tPBM therapy. Numerous studies have demonstrated that microstate disturbances in mental processes that are associated with neurological and psychiatric conditions, the most prominent pathology studied using the microstate approach is schizophrenia [46]. In schizophrenia, MS C is shorter in mean duration than healthy controls[62], and neurofeedback can be effectively employed to up-regulate the duration of MS C in patients with schizophrenia[63]. The microstate approach has also been used to investigate stroke, and the MS C parameters were lower represented in right damage [64]. These may suggest the potential neural mechanisms of 980 nm PW mode-based tPBM applied to the right prefrontal region in successfully treating neuropsychiatric disorders. But the alteration of EEG microstate seen in this study resting-state model in healthy populations may not be exactly replicated in clinical populations.

This study has several limitations. First, this study was limited to resting-state EEG and did not explore the association between behavior and brain function. Second, long-term follow-up interventions are often required in the application of neuropsychiatric disorders, but this study was limited to an experimental design with three times interventions and did not provide the response of EEG microstate following long-term intervention. To enhance the comprehensiveness of the neural mechanism of the tPBM, future research should expand the scope of examination to include a broader range of multiple tPBM sessions to explore the enhancement of cognitive functions and modulation of the brain functional networks.

In conclusion, the present study elucidates the microstate-specific neuromodulatory effects of PW mode-based tPBM therapy in 29 young participants. Compared to other non-invasive brain stimulation (NIBS) techniques, which may elicit sensations such as tingling or generate acoustic noise—especially the occurrence of butterfly-shaped hallucinations induced by electrical stimulation in the prefrontal region—tPBM therapy exhibits higher patient acceptability, consequently enhancing treatment adherence. Future works could make an effort to integrate tPBM delivery with EEG microstate analysis, which may further clarify the mechanisms of the tPBM-induced cognitive enhancements.

**Disclosures**

The authors declare that there are no financial interests, commercial affiliations, or other potential conflicts of interest that could have influenced the objectivity of this research or the writing of this paper

**Code and Data**

The corresponding author will provide the raw data and code supporting the study's conclusions upon reasonable request.

**Acknowledgments**

This project has been made possible with the financial support through the National Natural Science Foundation of China under grant 62275210; the National Leading Talent Program; the National Young Talent Program; the Key Research and Development Program of Shaanxi under grant 2022GY-313; Qinchuangyuan Scientists & Engineers of Shaanxi; the Xi'an Science and Technology Project under grant 23ZDCYJSGG0026-2023; the Fundamental Research Funds for the Central Universities under grant ZYTS23192, XJSJ24082, QTZX24079; the Xidian University Specially Funded Project for Interdisciplinary Exploration under grant TZJH2024036; the Innovation Fund of Xidian University. We would like to thank Wuhan Jin Laser Medical Technology Co., Ltd., particularly to Product Director Pan Yang, for providing 980 nm laser (Model Aurora-A3), whose exceptional precision and reliability significantly enhanced our experimental efficiency and data quality. We would like to express our sincere gratitude to DeepSeek v2.3 and Grammarly, which were used for language polishing to enhance the clarity and coherence of the manuscript. After using DeepSeek v2.3 and Grammarly, the authors reviewed and edited the content as needed, taking full responsibility for the content of the published article.

**CRediT authorship contribution statement**

**J.H.:** Writing – original draft, Validation, Methodology, Investigation, Formal analysis, Data curation, Conceptualization. **H.X.:** Writing – review & editing, Validation, Supervision, Resources, Project administration, Methodology, Investigation, Funding acquisition, Conceptualization. **Y.Y.:** Writing – original draft, Visualization, Software, Methodology, Formal analysis. **C.J.:** Data curation, Formal analysis, Visualization. **D.L.:** Methodology, Formal analysis. **L.Z.:** Methodology, Formal analysis. **X.W:** Data curation, Formal analysis. **L.T.:** Data curation, Formal analysis. **Z.D.:** Validation. **H.Y.:** Validation. **Z.Y.:** Writing – review & editing, Resources, Project administration. **M.J.:** Writing – review & editing, Resources, Project administration. **X.C.:** Writing – review & editing, Validation, Supervision, Resources, Project administration, Methodology, Investigation, Funding acquisition, Conceptualization. All authors reviewed the manuscript.

## References

[1] Hamblin, M R. Shining light on the head: Photobiomodulation for brain disorders. BBA Clin 2016;6113–124. https://doi.org/10.1016/j.bbacli.2016.09.002

[2] Al-Watban, F A H, X Y Zhang. The Comparison of Effects between Pulsed and CW Lasers on Wound Healing. J Clin Laser Med Surg 2004;24(1):15–18. https://doi.org/10.1089/104454704773660921

[3] Braverman, B, R J McCarthy, A D Ivankovich, D E Forde, M Overfield, M S Bapna. Effect of

helium-neon and infrared laser irradiation on wound healing in rabbits. Lasers Surg Med 1989;9(1):50–58. https://doi.org/10.1002/lsm.1900090111

[4] Sushko, B S, I P Lymans'kyĭ, S O Huliar. Action of the red and infrared electromagnetic waves of light-emitting diodes on the behavioral manifestation of somatic pain. Fiziolohichnyi Zhurnal 2007;53(3):51–60.

[5] Lampl, Y, J A Zivin, M Fisher, R Lew, L Welin, B Dahlof, P Borenstein, B Andersson, J Perez, C Caparo, S Ilic, U Oron. Infrared Laser Therapy for Ischemic Stroke: A New Treatment Strategy. Stroke 2007;38(6):1843–1849. https://doi.org/10.1161/STROKEAHA.106.478230

[6] Zivin, J A, G W Albers, N Bornstein, T Chippendale, B Dahlof, T Devlin, M Fisher, W Hacke, W Holt, S Ilic, S Kasner, R Lew, M Nash, J Perez, M Rymer, P Schellinger, D Schneider, S Schwab, R Veltkamp, M Walker, J Streeter. Effectiveness and Safety of Transcranial Laser Therapy for Acute Ischemic Stroke. Stroke 2009;40(4):1359–1364. https://doi.org/10.1161/STROKEAHA.109.547547

[7] Hacke, W, P D Schellinger, G W Albers, N M Bornstein, B L Dahlof, R Fulton, S E Kasner, A Shuaib, S P Richieri, S G Dilly, J Zivin, K R Lees, J Broderick, A Ivanova, K Johnston, B Norrving, A Alexandrov, D Brown, P Capone, D Chiu, W Clark, J Cochran, C Deredyn, T Devlin, W Hickling, G Howell, D Huang, S Hussain, S Mallenbaum, M Moonis, M Nash, M Rymer, R Taylor, M Tremwel, B Buck, J Perez, C Gerloff, B Greiwing, M Grond, G Hamman, T Haarmeiter, S Jander, M Köhrmann, M Ritter, D Schneider, J Sobesky, T Steiner, H Steinmetz, R Veltkamp, C Weimar, F Gruber, B Andersson, L Welin, D Leys, T Tatlisumak, A Luft, P Lyrer, P Michel, C Molina, T Segura. Transcranial Laser Therapy in Acute Stroke Treatment. Stroke 2014;45(11):3187–3193. https://doi.org/10.1161/STROKEAHA.114.005795

[8] Disner, S G, C G Beevers, F Gonzalez-Lima. Transcranial Laser Stimulation as Neuroenhancement for Attention Bias Modification in Adults with Elevated Depression Symptoms. Brain Stimulat 2016;9(5):780–787. https://doi.org/10.1016/j.brs.2016.05.009

[9] Sakalauskaitė, L, L S Hansen, J M Dubois, M P Larsen, G M Feijóo, M S Carstensen, K W Miskowiak, M Nguyen, L K H Clemmensen, P M Petersen, K Martiny. Rationale and design of a double-blinded, randomized placebo-controlled trial of 40 Hz light neurostimulation therapy for depression (FELIX). Ann Med 2024;56(1):2354852. https://doi.org/10.1080/07853890.2024.2354852

[10] Iosifescu, D V, K A Collins, A Hurtado-Puerto, M K Irvin, J A Clancy, A M Sparpana, E F Sullivan, Z Parincu, E-M Ratai, C J Funes, A Weerasekera, J P Dmochowski, P Cassano. Grant Report on the Transcranial near Infrared Radiation and Cerebral Blood Flow in Depression (TRIADE) Study. Photonics 2023;10(1):90. https://doi.org/10.3390/photonics10010090

[11] Berman, M H. Shining new light on dementia treatment: ADAS-Cog and quantitative EEG evidence of cognitive improvement in early and mid-stage dementia using self-administered 1070 nm transcranial near-infrared photobiomodulation. Alzheimers Dement 2021;17 (Suppl. 8)(S8):e057541–e057541. https://doi.org/10.1002/alz.057541

[12] Tao, L, Q Liu, F Zhang, Y Fu, X Zhu, X Weng, H Han, Y Huang, Y Suo, L Chen, X Gao, X Wei. Microglia modulation with 1070-nm light attenuates Aβ burden and cognitive impairment in Alzheimer's disease mouse model. Light Sci Appl 2021;10(1):179.

https://doi.org/10.1038/s41377-021-00617-3

[13] Freitas, L F de, M R Hamblin. Proposed Mechanisms of Photobiomodulation or Low-Level Light Therapy. IEEE J Sel Top Quantum Electron 2016;22(3):348–364. https://doi.org/10.1109/JSTQE.2016.2561201

[14] Salehpour, F, J Mahmoudi, F Kamari, S Sadigh-Eteghad, S H Rasta, M R Hamblin. Brain Photobiomodulation Therapy: a Narrative Review. Mol Neurobiol 2018;55(8):6601–6636. https://doi.org/10.1007/s12035-017-0852-4

[15] Lee, T, Z Ding, A S Chan. Can transcranial photobiomodulation improve cognitive function? A systematic review of human studies. Ageing Res Rev 2023;83101786. https://doi.org/10.1016/j.arr.2022.101786

[16] Fernandes, F, S Oliveira, F Monteiro, M Gasik, F S Silva, N Sousa, Ó Carvalho, S O Catarino. Devices used for photobiomodulation of the brain—a comprehensive and systematic review. J NeuroEngineering Rehabil 2024;21(1):53. https://doi.org/10.1186/s12984-024-01351-8

[17] Plaksin, M. Thermal Transients Excite Neurons through Universal Intramembrane Mechanoelectrical Effects. Phys Rev X 2018;8(1) https://doi.org/10.1103/PhysRevX.8.011043

[18] Gigo-Benato, D, S Geuna, A de Castro Rodrigues, P Tos, M Fornaro, E Boux, B Battiston, M G Giacobini-Robecchi. Low-power laser biostimulation enhances nerve repair after end-to-side neurorrhaphy: a double-blind randomized study in the rat median nerve model. Lasers Med Sci 2004;19(1):57–65. https://doi.org/10.1007/s10103-004-0300-3

[19] El Sayed, S O, M Dyson. Effect of laser pulse repetition rate and pulse duration on mast cell number and degranulation. Lasers Surg Med 1996;19(4):433–437. https://doi.org/10.1002/(SICI)1096-9101(1996)19:4<433::AID-LSM8>3.0.CO;2-T

[20] Ueda, Y, N Shimizu. Effects of Pulse Frequency of Low-Level Laser Therapy (LLLT) on Bone Nodule Formation in Rat Calvarial Cells. J Clin Laser Med Surg 2004;21(5):271–277. https://doi.org/10.1089/104454703322564479

[21] Lapchak, P A, K F Salgado, C H Chao, J A Zivin. Transcranial near-infrared light therapy improves motor function following embolic strokes in rabbits: An extended therapeutic window study using continuous and pulse frequency delivery modes. Neuroscience 2007;148(4):907–914. https://doi.org/10.1016/j.neuroscience.2007.07.002

[22] Lapchak, P A, L De Taboada. Transcranial near infrared laser treatment (NILT) increases cortical adenosine-5′-triphosphate (ATP) content following embolic strokes in rabbits. Brain Res 2010;1306100–105. https://doi.org/10.1016/j.brainres.2009.10.022

[23] Ando, T, W Xuan, T Xu, T Dai, S K Sharma, G B Kharkwal, Y-Y Huang, Q Wu, M J Whalen, S Sato, M Obara, M R Hamblin. Comparison of Therapeutic Effects between Pulsed and Continuous Wave 810-nm Wavelength Laser Irradiation for Traumatic Brain Injury in Mice. PLOS One 2011;6(10):e26212. https://doi.org/10.1371/journal.pone.0026212

[24] Thunshelle, C, M R Hamblin. Transcranial Low-Level Laser (Light) Therapy for Brain Injury. Photomed Laser Surg 2016;34(12):587–598. https://doi.org/10.1089/pho.2015.4051

[25] Tang, L, H Jiang, M Sun, M Liu. Pulsed transcranial photobiomodulation generates distinct beneficial neurocognitive effects compared with continuous wave transcranial light. Lasers Med Sci 2023;38(1):203. https://doi.org/10.1007/s10103-023-03865-4

[26] Zhang, X, X Miao, H Jiang, Y Ren, L Huo, M Liu, H Chen. Advanced Intervention Effects of

Pulsed and Steady Transcranial Photobiomodulation on Sleep, Mood, and EEG Signal Regulation. J Biophotonics 2025;e70004. https://doi.org/10.1002/jbio.70004

[27] Weerasekera, A, D R A Coelho, E-M Ratai, K A Collins, A M H Puerto, L De Taboada, M B Gersten, J A Clancy, M J Hoptman, M K Irvin, A M Sparpana, E F Sullivan, X Song, A Adib, P Cassano, D V Iosifescu. Dose-dependent effects of transcranial photobiomodulation on brain temperature in patients with major depressive disorder: a spectroscopy study. Lasers Med Sci 2024;39(1):249. https://doi.org/10.1007/s10103-024-04198-6

[28] Caldieraro, M A, T Laufer-Silva, P Cassano. Dosimetry and Clinical Efficacy of Transcranial Photobiomodulation for Major Depression Disorder: Could they Guide Dosimetry for Alzheimer’s Disease? J Alzheimers Dis 2021;83(4):1453–1469. https://doi.org/10.3233/JAD-210586

[29] Dong, S, R Zhang, J Xue, Y Suo, X Wei. Quantitative simulation of near-infrared light treatment for Alzheimer’s disease using patient-individualized optical-parametric phantoms. Neurophotonics 2025;12(1):015010. https://doi.org/10.1117/1.NPh.12.1.015010

[30] Salehpour, F, M R Hamblin, J O DiDuro. Rapid Reversal of Cognitive Decline, Olfactory Dysfunction, and Quality of Life Using Multi-Modality Photobiomodulation Therapy: Case Report. Photobiomodulation Photomed Laser Surg 2019;37(3):159–167. https://doi.org/10.1089/photob.2018.4569

[31] Kang, S, L Li, S Shahdadian, A Wu, H Liu. Site- and electroencephalogram-frequency–specific effects of 800-nm prefrontal transcranial photobiomodulation on electroencephalogram global network topology in young adults. Neurophotonics 2025;12(1):015011. https://doi.org/10.1117/1.NPh.12.1.015011

[32] Shahdadian, S, X Wang, H Wanniarachchi, A Chaudhari, N C D Truong, H Liu. Neuromodulation of brain power topography and network topology by prefrontal transcranial photobiomodulation. J Neural Eng 2022;19(6):066013. https://doi.org/10.1088/1741-2552/ac9ede

[33] Truong, N C D, X Wang, H Liu. Temporal and spectral analyses of EEG microstate reveals neural effects of transcranial photobiomodulation on the resting brain. Front Neurosci 2023;171247290. https://doi.org/10.3389/fnins.2023.1247290

[34] Etemadi, A, S Sadatmansouri, F Sodeif, F Jalalishirazi, N Chiniforush. Photobiomodulation Effect of Different Diode Wavelengths on the Proliferation of Human Gingival Fibroblast Cells. Photochem Photobiol 2021;97(5):1123–1128. https://doi.org/10.1111/php.13463

[35] Kobiela Ketz, A, K R Byrnes, N E Grunberg, C E Kasper, L Osborne, B Pryor, N L Tosini, X Wu, J J Anders. Characterization of Macrophage/Microglial Activation and Effect of Photobiomodulation in the Spared Nerve Injury Model of Neuropathic Pain. Pain Med 2017;18(5):932–946. https://doi.org/10.1093/pm/pnw144

[36] Anders, J J, H Moges, X Wu, I D Erbele, S L Alberico, E K Saidu, J T Smith, B A Pryor. In vitro and in vivo optimization of infrared laser treatment for injured peripheral nerves. Lasers Surg Med 2014;46(1):34–45. https://doi.org/10.1002/lsm.22212

[37] Masoumipoor, M, S B Jameie, A Janzadeh, F Nasirinezhad, M Soleimani, M Kerdary. Effects of 660- and 980-nm low-level laser therapy on neuropathic pain relief following chronic constriction injury in rat sciatic nerve. Lasers Med Sci 2014;29(5):1593–1598. https://doi.org/10.1007/s10103-014-1552-1

[38] Hakimiha, N, M M Dehghan, H Manaheji, J Zaringhalam, S Farzad-Mohajeri, R Fekrazad, N Moslemi. Recovery of inferior alveolar nerve by photobiomodulation therapy using two laser wavelengths: A behavioral and immunological study in rat. J Photochem Photobiol B 2020;204111785. https://doi.org/10.1016/j.jphotobiol.2020.111785

[39] Morries, L D, P Cassano, T A Henderson. Treatments for traumatic brain injury with emphasis on transcranial near-infrared laser phototherapy. Neuropsychiatr Dis Treat 2015;112159–2175. https://doi.org/10.2147/NDT.S65809

[40] Cassano, P, A P Tran, H Katnani, B S Bleier, M R Hamblin, Y Yuan, Q Fang. Selective photobiomodulation for emotion regulation: model-based dosimetry study. Neurophotonics 2019;6(1):015004. https://doi.org/10.1117/1.NPh.6.1.015004

[41] Yuan, Y, P Cassano, M Pias, Q Fang. Transcranial photobiomodulation with near-infrared light from childhood to elderliness: simulation of dosimetry. Neurophotonics 2020;7(1):015009. https://doi.org/10.1117/1.NPh.7.1.015009

[42] Fan, L, H Li, J Zhuo, Y Zhang, J Wang, L Chen, Z Yang, C Chu, S Xie, A R Laird, P T Fox, S B Eickhoff, C Yu, T Jiang. The Human Brainnetome Atlas: A New Brain Atlas Based on Connectional Architecture. Cereb Cortex 2016;263508–3526. https://doi.org/10.1093/cercor/bhw157

[43] Li, J, N Li, X Shao, J Chen, Y Hao, X Li, B Hu. Altered Brain Dynamics and Their Ability for Major Depression Detection Using EEG Microstates Analysis. IEEE Trans Affect Comput 2023;14(3):2116–2126. https://doi.org/10.1109/TAFFC.2021.3139104

[44] Gold, M C, S Yuan, E Tirrell, E F Kronenberg, J W D Kang, L Hindley, M Sherif, J C Brown, L L Carpenter. Large-scale EEG neural network changes in response to therapeutic TMS. Brain Stimulat 2022;15(2):316–325. https://doi.org/10.1016/j.brs.2022.01.007

[45] Dong, L, F Li, Q Liu, X Wen, Y Lai, P Xu, D Yao. MATLAB Toolboxes for Reference Electrode Standardization Technique (REST) of Scalp EEG. Front Neurosci 2017;11 https://doi.org/10.3389/fnins.2017.00601

[46] Michel, C M, T Koenig. EEG microstates as a tool for studying the temporal dynamics of whole-brain neuronal networks: A review. NeuroImage 2018;180577–593. https://doi.org/10.1016/j.neuroimage.2017.11.062

[47] Britz, J, D Van De Ville, C M Michel. BOLD correlates of EEG topography reveal rapid resting-state network dynamics. NeuroImage 2010;52(4):1162–1170. https://doi.org/10.1016/j.neuroimage.2010.02.052

[48] Férat, V, M Scheltienne, D Brunet, T Ros, C Michel. Pycrostates: a Python library to study EEG microstates. J Open Source Softw 2022;7(78):4564. https://doi.org/10.21105/joss.04564

[49] Chu, C, Z Zhang, J Wang, Z Li, X Shen, X Han, L Bai, C Liu, X Zhu. Temporal and spatial variability of dynamic microstate brain network in early Parkinson's disease. Npj Park Dis 2023;9(1):1–12. https://doi.org/10.1038/s41531-023-00498-w

[50] Imperatori, L S, M Betta, L Cecchetti, A Canales-Johnson, E Ricciardi, F Siclari, P Pietrini, S Chennu, G Bernardi. EEG functional connectivity metrics wPLI and wSMI account for distinct types of brain functional interactions. Sci Rep 2019;9(1):8894. https://doi.org/10.1038/s41598-019-45289-7

[51] Vinck, M, R Oostenveld, M van Wingerden, F Battaglia, C M A Pennartz. An improved index of phase-synchronization for electrophysiological data in the presence of volume-

conduction, noise and sample-size bias. NeuroImage 2011;55(4):1548–1565. https://doi.org/10.1016/j.neuroimage.2011.01.055

[52] Farahani, F V, W Karwowski, N R Lighthall. Application of Graph Theory for Identifying Connectivity Patterns in Human Brain Networks: A Systematic Review. Front Neurosci 2019;13 https://doi.org/10.3389/fnins.2019.00585

[53] Boccaletti, S, V Latora, Y Moreno, M Chavez, D-U Hwang. Complex networks: Structure and dynamics. Phys Rep 2006;424(4):175–308. https://doi.org/10.1016/j.physrep.2005.10.009

[54] Redcay, E, J M Moran, P L Mavros, H Tager-Flusberg, J D E Gabrieli, S Whitfield-Gabrieli. Intrinsic functional network organization in high-functioning adolescents with autism spectrum disorder. Front Hum Neurosci 2013;7 https://doi.org/10.3389/fnhum.2013.00573

[55] Gao, W, J H Gilmore, K S Giovanello, J K Smith, D Shen, H Zhu, W Lin. Temporal and Spatial Evolution of Brain Network Topology during the First Two Years of Life. PLOS One 2011;6(9):e25278. https://doi.org/10.1371/journal.pone.0025278

[56] Khare, S K, U R Acharya. An explainable and interpretable model for attention deficit hyperactivity disorder in children using EEG signals. Comput Biol Med 2023;155106676. https://doi.org/10.1016/j.compbiomed.2023.106676

[57] Lundberg, S M, S-I Lee. A Unified Approach to Interpreting Model Predictions. 31st Conf Neural Inf Process Syst NIPS 2017;4768–4777.

[58] Feng, R, J Yang, H Huang, Z Chen, R Feng, N U Farrukh Hameed, X Zhang, J Hu, L Chen, S Lu. Spatiotemporal Microstate Dynamics of Spike-free Scalp EEG Offer a Potential Biomarker for Refractory Temporal Lobe Epilepsy. IEEE Trans Med Imaging 2024;1–1. https://doi.org/10.1109/TMI.2024.3453377

[59] Lamoš, M, M Bočková, S Goldemundová, M Baláž, J Chrastina, I Rektor. The effect of deep brain stimulation in Parkinson's disease reflected in EEG microstates. Npj Park Dis 2023;9(1):1–7. https://doi.org/10.1038/s41531-023-00508-x

[60] Ghaderi, A H, A Jahan, F Akrami, M Moghadam Salimi. Transcranial photobiomodulation changes topology, synchronizability, and complexity of resting state brain networks. J Neural Eng 2021;18(4):046048. https://doi.org/10.1088/1741-2552/abf97c

[61] Zomorrodi, R, G Loheswaran, A Pushparaj, L Lim. Pulsed Near Infrared Transcranial and Intranasal Photobiomodulation Significantly Modulates Neural Oscillations: a pilot exploratory study. Sci Rep 2019;9(1):6309. https://doi.org/10.1038/s41598-019-42693-x

[62] Rieger, K, L Diaz Hernandez, A Baenninger, T Koenig. 15 Years of Microstate Research in Schizophrenia – Where Are We? A Meta-Analysis. Front Psychiatry 2016;7 https://doi.org/10.3389/fpsyt.2016.00022

[63] Diaz Hernandez, L, K Rieger, A Baenninger, D Brandeis, T Koenig. Towards Using Microstate-Neurofeedback for the Treatment of Psychotic Symptoms in Schizophrenia. A Feasibility Study in Healthy Participants. Brain Topogr 2016;29(2):308–321. https://doi.org/10.1007/s10548-015-0460-4

[64] Zappasodi, F, P Croce, A Giordani, G Assenza, N M Giannantoni, P Profice, G Granata, P M Rossini, F Tecchio. Prognostic Value of EEG Microstates in Acute Stroke. Brain Topogr 2017;30(5):698–710. https://doi.org/10.1007/s10548-017-0572-0